\documentclass[11pt,twocolumn]{article}

\usepackage{times}
\usepackage{overcite}

\begin{document}

\twocolumn[
\begin{@twocolumnfalse}
{
\begin{center}
{\bf\Large Reply to Comment by Alexandrov and Bratkovsky}\\*[0.1cm]
{\bf\Large [cond-mat/0603467, cond-mat/0606366]}\\*[0.3cm]
{\bf\large Michael~Galperin${}^1$, Abraham~Nitzan${}^2$, and Mark~A.~Ratner${}^1$}\\*[0.2cm]
{\small\it ${}^1$Department of Chemistry and Nanotechnology Center,
           Northwestern University, Evanston, IL 60208}\\
{\small\it ${}^2$School of Chemistry, The Sackler Faculty of Science,
           Tel Aviv University, Tel Aviv 69978, Israel}\\*[0.2cm]
\end{center}
{\small
The submitted Comment is based on using an isolated quantum dot approach
to discuss the situation where the coupling to the leads is considerable
(not negligible). This finite lead coupling
is the situation in most molecular transport
junctions. In such situations the population on the molecule is not 
static and dynamical effects (fluctuations) are crucial.
While the discussion in Ref.~\citeonline{AlexandrovBratkovsky_3}, 
and in the Comment may well apply for an isolated molecule, 
it is simply irrelevant for the very 
different transport situation analyzed in our original contribution.\\*[0.5cm]
}
}
\end{@twocolumnfalse}
]

In their preceding comment on our earlier paper\cite{hysteresis}, 
Alexandrov and Bratkovsky (AB) bring up the following claims:
\begin{enumerate}
\item \label{issue1}
      A previous work by AB\cite{AlexandrovBratkovsky_3} that discusses 
      bistability in correlated transport through a degenerate molecular 
      quantum dot was ``surprisingly'' not cited in our paper.\cite{hysteresis}
\item \label{issue2}
      The mean field approximation used in Ref.~\citeonline{hysteresis} is 
      invalid.
\item \label{issue3}
      The bistability and hysteresis phenomenon discussed in 
      Ref.~\citeonline{hysteresis} is an artifact of this invalid mean field 
      approximation.
\end{enumerate}

Indeed, while we have cited the paper on bistable tunneling current by 
AB,\cite{AlexandrovBratkovsky_2} we indeed overlooked the 
paper~\citeonline{AlexandrovBratkovsky_3} by these authors. 
In retrospect we believe (for reasons given below) that the paper is only 
marginally relevant to our 
work, however discussing it at the time would have make some issues clearer 
and could have saved the present exchange, so we regret our failing to do so. 
In what follows we address issues (\ref{issue2}) and (\ref{issue3}).

Consider first the mean field approximation. Let two coupled subsystems 
$a$ and $b$ of system $s$ be described by the Hamiltonian $H_s=H_a+H_b+V_{ab}$ 
and let system $a$ be characterized by two states, $|1_a>$ and $|2_a>$. 
Assume furthermore that an additional process causes a rapid interchange 
between states $|1_a>$ and $|2_a>$, and that this process by itself would 
bring subsystem $a$ to equilibrium with probability $P_1$ to be in state 
$|1_a>$  and $P_2$ to be in state $|2_a>$. How good is an approximation by 
which the dynamics of system $b$ is assumed to be governed by the Hamiltonian 
$H_b+<V_{ab}>_a$ where $<\ldots>_a$ denotes an average over the state of $a$, 
and in particular when can $<V_{ab}>_a$ be represented by 
$P_1<1_a|V_{ab}|1_a>+P_2<2_a|V_{ab}|2_a>$?

Describing the motion of subsystem $b$ by the Hamiltonian $H_b+<V_{ab}>_a$ 
(while deriving the dynamics of $a$ from the frozen instantaneous state of $b$)
is the essence of the adiabatic (Born-Oppenheimer) approximation used in 
quantum
molecular dynamics, and relies on the assumption that system $a$ is much faster
than system $b$. The correct identification of $<V_{ab}>_a$ depends on details 
of this assumption. In particular, in the example constructed above, 
identifying $<V_{ab}>_a$ as given by $P_1<1_a|V_{ab}|1_a>+P_2<2_a|V_{ab}|2_a>$ 
(rather than, e.g. moving system $b$ under the potential $<1_a|V_{ab}|1_a>$ 
when $a$ is in state $|1_a>$ and under $<2_a|V_{ab}|2_a>$ when $a$ is in state
$|2_a>$) is valid {\em provided that the dynamics of the 
$|1_a>\leftrightarrow |2_a>$ is fast on the timescale that characterizes the 
dynamics of system $b$.}

In the present problem system $b$ is the vibration of frequency $\omega_0$ and 
system $a$ is the molecule, which is characterized by two states -– 
occupied and unoccupied. 
The process that interchanges between the two states is the 
molecule-leads electron exchange, whose characteristic timescale is 
$\Gamma^{-1}=2\pi V^2\rho$ where $V$ is the molecule-lead coupling and $\rho$ 
-– the electronic density of states in the lead. From the above discussion 
it is clear that the condition for the validity of the mean-field approximation 
as applied in Ref.~\citeonline{hysteresis} is $\Gamma>\hbar\omega_0$. 
In molecular 
junctions with chemical binding between the electrodes and the molecular bridge
$\Gamma$ is of the order $0.1-1$~eV while typical vibrational frequencies span 
the range $0.001-0.3$~eV. There is thus a wide regime in which this 
approximation is valid.

In the preceding comment, AB illustrate the failure of the mean field 
approximation for a model in which $\Gamma=0$. Clearly this illustration 
is irrelevant for the problem at hand. They then proceed to discuss, under 
what they call ``the correct procedure'', a different model 
(a recast of Ref.~\citeonline{AlexandrovBratkovsky_3}) in which bistability 
stems from electron-electron interaction in a d-fold degenerate molecule. 
Again, analysis of this model is done by neglecting the molecule-leads coupling
and, furthermore, disregarding the effect of the electron-vibration interaction 
on the state of the vibration -– two principal ingredients of the model of 
Ref.~\citeonline{hysteresis}. While we have no objections to this model and its 
consequences, it can hardly be regarded as a ``correct procedure'' with 
respect to our model, or, indeed, to the molecular junction problem that
we analyzed. We note that similar models, with similar conclusions 
about the inherent bistability in the system (though with different views about
its consequence –- see below) have been recently discussed by several 
authors.\cite{GogolinKomnik,MitraAleinerMillis_prb,MitraAleinerMillis_prl,MozyrskyHastingsMartin}

Next consider the consequences of the bistable nature of the nuclear potential 
surface obtained in the mean field approximation. Obviously, as emphasized by 
Mitra~et~al,\cite{MitraAleinerMillis_prl} this bistability does not imply any 
phase transition property. The observation of hysteresis or, instead, 
switching behavior (telegraphic noise) depends on the timescale of changing 
the external control parameter (e.g. the bias or the gate potential) relative 
to the rates of transitions between locally stable 
states.\cite{MozyrskyHastingsMartin} Both phenomena are observed in molecular 
junctions, and the question whether they are accounted for by the present 
mechanism should be settled by determining these rates. This issue has not 
been raised or discussed in the preceding comment by AB.  

We end our reply with several comments:

\noindent
1. For an isolated molecule (MQD in the AB language) the molecular level is 
well defined, and is shifted from $\varepsilon_0$ to 
$\varepsilon_0-M^2/\omega_0$ by the electron-vibration interaction 
(reorganization energy). Coupling to the leads gives rise to electronic level 
population that fluctuates on the characteristic timescale $\Gamma^{-1}$. 
The observation that for $\Gamma>\hbar\omega_0$ the electronic energy shift 
does depend on the average population of molecular level, was first made in
Refs.~\citeonline{HewsonNewns_1,HewsonNewns_2}. 
The isolated MQD consideration of~\citeonline{AlexandrovBratkovsky_3} 
misses the physics of this situation, 
which is physically relevant to molecular transport junctions.

\noindent
2. The AB Comment refers to the approximation $\hat n^2=n_0\hat n$ as 
``a spurious self-interaction of a single polaron with itself''. In fact, 
this term represents the interaction of a tunneling electron at a particular 
time with the polarization charge established in response to electrons 
(many tunneling events) which traversed the junction at earlier times. 
This ``self-interaction'' is legitimate in the (physically relevant) case 
where coupling to the leads (disregarded in the AB treatment) is taken into 
account. 

\noindent
3. The treatment of Ref.~\citeonline{MozyrskyHastingsMartin} of the same 
one-level model, which goes beyond the mean-field approximation, also leads to 
multistability in agreement with our treatment. The statement in 
Ref.~\citeonline{AlexandrovBratkovsky_3} that ``the Born-Oppenheimer 
approximation does not apply to nondegenerate level, since there are no fast 
(compared to the characteristic phonon time $1/\omega_0$) electron 
transitions within the dot'' holds for the isolated MQD. 
In molecular wires the molecule-metal 
electron exchange with timescale $\Gamma^{-1}$ provides this fast process. 

\noindent
4. Contrary to statements made in the AB Comment, taking into account the 
coupling with the leads (after the Lang-Firsov transformation applied to the 
Hamiltonian as is done in Ref.~\citeonline{AlexandrovBratkovsky_3}) does 
provide nonlinearity, since the effective couplings do depend on the electron 
population on the bridge. Confusion here is due to the approximation made 
in~\citeonline{AlexandrovBratkovsky_3} which neglects the effect of the 
electron-vibration coupling on the vibration state. 
For detailed discussion see~\citeonline{strong_elph}.

In conclusion, while the discussion in Ref.~\citeonline{AlexandrovBratkovsky_3}, 
and in the AB Comment apply for an isolated molecule, it is simply
irrelevant for the very different transport situation analyzed in our 
original contribution.


\begin{thebibliography}{99}
\bibitem{hysteresis}M.~Galperin, M.~A.~Ratner, and A.~Nitzan, 
   Nano~Lett. \textbf{5}, 125 (2005).
\bibitem{AlexandrovBratkovsky_3}A.~S.~Alexandrov and A.~M.~Bratkovsky,
   Phys.~Rev.~B \textbf{67}, 235312 (2003).
\bibitem{AlexandrovBratkovsky_2}A.~S.~Alexandrov, A.~M.~Bratkovsky,
   and R.~S.~Williams, Phys.~Rev.~B \textbf{67}, 075301 (2003).
\bibitem{GogolinKomnik} A.~Gogolin and A.~Komnik, cond-mat/0207513 (2002).
\bibitem{MitraAleinerMillis_prb}A.~Mitra, I.~Aleiner, and A.~J.~Millis,
   Phys.~Rev.~B \textbf{69}, 245302 (2004).
\bibitem{MitraAleinerMillis_prl}A.~Mitra, I.~Aleiner, and A.~J.~Millis,
   Phys.~Rev.~Lett. \textbf{94}, 076404 (2005).
\bibitem{MozyrskyHastingsMartin}D.~Mozyrsky, M.~B.~Hastings, and I.~Martin,
   Phys.~Rev.~B \textbf{73}, 035104 (2006).
\bibitem{HewsonNewns_1}A.~C.~Hewson and D.~M.~Newns, 
   Proc. 2nd Int. Conf. Solid Surfaces, Japan.~J.~Appl.~Phys. Suppl.~2, Pt.~2,
   121 (1974).
\bibitem{HewsonNewns_2}A.~C.~Hewson and D.~M.~Newns,
   J.~Phys.~C: Solid~State~Phys. \textbf{12}, 1665 (1979).
\bibitem{strong_elph}M.~Galperin, A.~Nitzan, and M.~A.~Ratner,
   Phys.~Rev.~B \textbf{73}, 045314 (2006).
\end{thebibliography}
\end{document}